# Design and Fabrication Of Multiplexed One-Port SAW Resonators On A Single Chip

A.Westrelin[1*], N. Abdallah[1]P, P. Debavelear[2], M. Lefevbre[2], K. Lmimouni[1], O.Stienne[2], B. Hafsi[1,2*]

[1] Univ. Lille, CNRS, Centrale Lille, Univ. Polytechnique Hauts-de-France, UMR 8520 - IEMN, F-59000 Lille, France.
[2] Icam School of Engineering, 06 rue Auber, 59800, Lille, France
 Corresponding author:  alexandre.wetrelin@iemn.fr, bilel.hafsi@icam.fr

**Abstract—** Shear horizontal surface acoustic wave (SH-SAW) sensors are considered as a viable option for label-free, sensitive, real-time, and cost-effective detection. In this paper, we present the design, the fabrication and the characterization of multiplexed SAW sensors based on a single chip SAW resonator. Operating near the industrial, scientific and medical (ISM) bands, resonators were microfabricated on 42.75° Y (ST) cut Quartz. This substrate was selected due to its traditional association with temperature stability in the design of SAW devices. Parameters like aperture, number of interdigitated electrode (IDT) pairs, IDT wavelengths, and number of reflectors were studied experimentally to enhance electrical responses. Resonance frequency, quality factor and electromechanical coupling were extracted through impedance measurements and fitted with a modified Butterworth-Van Dyke (MBVD) model. Based on these results, we propose a novel design for a multi-frequency device that can serve as a single electronic nose for complex gases detection.

## I. INTRODUCTION

The development of Surface Acoustic Wave (SAW) resonators based on various materials and designs has been extensively explored, particularly for applications such as filters [1], RFID tags [2], and sensors [3],[4]. In fact, the major manufacturing challenges have been overcome since those devices entered the industrial market two decades ago. Current research continues to concentrate on discovering new detection approach and expanding applications [5]. The fabrication of SAW resonators involves a complex process influenced by numerous factors that can significantly impact their performance. This presents a considerable challenge in developing arrays of devices that not only operate within defined frequency bands such as the 433 MHz range for industrial, scientific, and medical applications [6],[7] , but also maintain high quality factors and distinct, yet closely spaced resonance frequencies. These challenges are amplified by the sensitivity of these devices to even minor variations in the fabrication process.

Furthermore, there is a growing need to design multiplexed resonators that can provide multiple electrical responses for advanced applications like gas sensing. Such capabilities are essential for enhancing the functionality of electronic noses (e-noses) [8],[9], which rely on the ability to detect and discriminate among complex gas mixtures. A key feature of these systems is their wireless interrogation capability, which enables the remote sensing of hazardous environments [10]. Integrating multiple sensors into a single device for gas detection is crucial in enhancing the multi-detection paradigm, particularly for environments where the presence of diverse gases needs to be monitored simultaneously. This approach allows for more accurate and comprehensive monitoring of air quality [11], industrial emissions [12], or agriculture [13], where detecting multiple types of gases can be critical for safety and compliance.

This paper explores the potential of multiplexed SAW resonators to significantly improve the sensitivity and selectivity of e-nose responses, positioning them as critical components in the development of sophisticated sensing technologies. The capability of multi-sensing devices has been demonstrated in numerous studies for a variety of applications. Malocha DC et al [14] demonstrated a passive wireless multi-sensor based on YZ LiNbO$_3$ substrate and used for orthogonal frequency coding (OFC). These sensors are notably compact and rugged, integrating radio frequency identification (RFID) directly on the chip. They are engineered to operate across a wide range of conditions, from cryogenic to high temperatures. Bruckner, G et al. [15] investigated multiple SAW sensors based on reflective delay lines that allow the parallel readout of four independent temperature sensors in the 2.45 GHz ISM-band. A multi-frequency surface acoustic wave resonators (SWARs) based humidity sensor, operating at three distinct resonant frequencies—180, 200, and 250 MHz was reported in [16]. These resonators share a common sensing area, enabling straightforward evaluation of the sensor's performance across various operating frequencies. At the same time, a number of sensors based on multiplexed SAW devices have been proposed in a wide range of applications [17],[18],[19],[20], with an increased interest in gaz detection. However, none of them is focused on achieving operation using a one port multi-frequency detection on a single chip device.

In this study, we aimed to develop single-chip multiplexed SAW resonators based on an ST-cut quartz substrate. We assessed the precision of our fabrication process in their ability to produce devices with distinct frequency responses near the 434 MHz ISM band. Our approach involved analyzing the impact of fabrication parameters, such as aperture, number of interdigitated electrode (IDT) pairs, IDT wavelengths, and number of reflectors, on device performance. We evaluate the electrical performance through the fabrication of 216 resonators, all designed and distributed across a single wafer. Statistical analysis of these resonators allowed us to tune the resonant frequencies, the Q-factor and impedance matching. Based on these results, we used these optimized devices to propose an original design based on a multi-frequency single device intended to be used as a single electronic nose for gas detection.

## II. EPERIMENTAL PROCEDURE

### A. Design parameters

The structure of the Surface Acoustic Wave (SAW) resonator discussed in this article is illustrated in Figure **1**. It includes a set of reflector gratings, forming an acoustic cavity, and an Interdigital Transducer (IDT) which is responsible for coupling acoustic energy into the cavity. These components are fabricated on a piezoelectric substrate, essential for the operation of the IDT. Specifically, this study uses an ST-cut quartz substrate, with its material properties detailed in Table **1**.

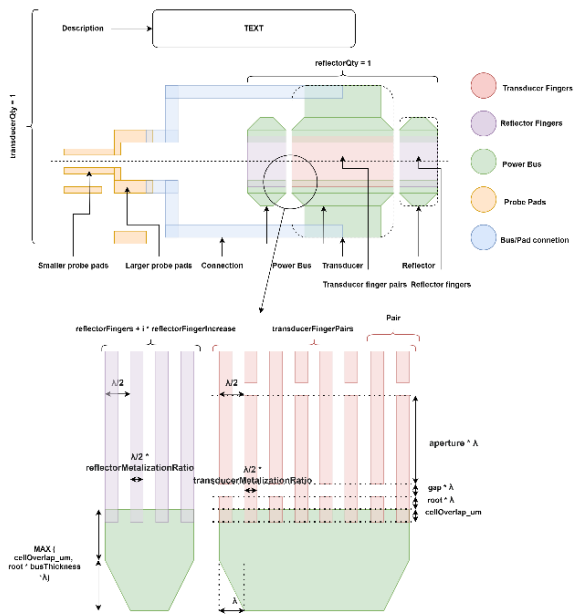

Fig 1. Design parameters of a one-port SAW resonator.

The surface acoustic wave is generated by applying an appropriate electrical signal to the Interdigital Transducer (IDT). The reflector gratings consist of a series of shorted metal strips, acting as mirrors within their stop band. They reflect the incoming waves, thereby creating a resonant cavity with an effective length $L_g$. The base resonance frequency of the resonator is determined by the velocity of the surface acoustic wave (v) and its central wavelength (λ) as follows [21]:

$$f = \frac{v}{\lambda} \quad (1)$$

Typical metals used for constructing the IDT include aluminum, gold, nickel, copper, chromium, platinum, and silver [22],[23]. Gold electrodes were fabricated to ensure electrochemical deposition methods of conducting polymers [24], as shown in our previous work. The metallic IDT finger width in the studied devices matches the quarter wavelength of the Surface Acoustic Wave (SAW).

The design of SAW devices was varied based on five parameters: transducer IDT pairs (Np), acoustic aperture (W), number of IDT pairs in reflector (Nr), IDT wavelengths (λ), and number of reflectors (Nbr), as shown in Table **1**. The electrode metallization ratio was kept constant. The design constraints were the response gain, working frequency ($f_r$) and a quality-factor (Q). The goal was to develop devices with 50 ohms matched impedance to simplify their integration with measurement equipment or RF front end systems.

Table 1. Design parameters of SAW devices

| **Construction parameters** |
| --- |
| Target Frequency = [430.5MHz; 438.7 MHz] |
| Metal Thickness: (Ti/Au)=(10nm/100nm) |
| Reflector Qty: Nbr=[40;100] |
| Transducer Fingers: NP=[10;30] |
| Reflector Fingers: Nr=[10;30] |
| Separation: Lg= λ/4 |
| Gap= 1 λ |
| Root= 4λ |
| Metallization ratio= 0.5 |
| **Material parameters** |
| Material: Quartz |
| Orientation: ST-Cut (+42.75Y) |
| Free velocity: V=5100 m.s$^{-1}$ |
| Wave velocity : v=4475 m.s$^{-1}$ |
| Prob_floting_ratio= 0.9 |

### B. Fabrication and characterization process

A 4-inch ST-cut quartz wafer with a single side polished (Precision Micro-Optics Inc.) was used as the piezoelectric substrate in this work. The wafer was cleaned by sonication in acetone, isopropanol, for 15 min, then by ultraviolet irradiation in an ozone atmosphere (ozonolysis) for 30 min to eliminate dust and contaminants from the substrate's surface. Following the cleaning process, the substrate was dried with high-purity nitrogen gas. The cleaned substrate was subsequently coated with a positive photoresist bilayer (COPO El 13% / PMMA). After the e-beam lithography process, the bilayer photoresist was developed in MIBK (methyl isobutyl ketone) and IPA (isopropyl alcohol) for 40s. A (Ti/Au) IDT electrodes, used to construct SAW devices, were then thermally evaporated on top of the substrate. Finally, a lift-off process was conducted by immersing in organic solvent (SVC-14) at 70°C for 2h, followed by rinsing in acetone and IPA, and drying with nitrogen.

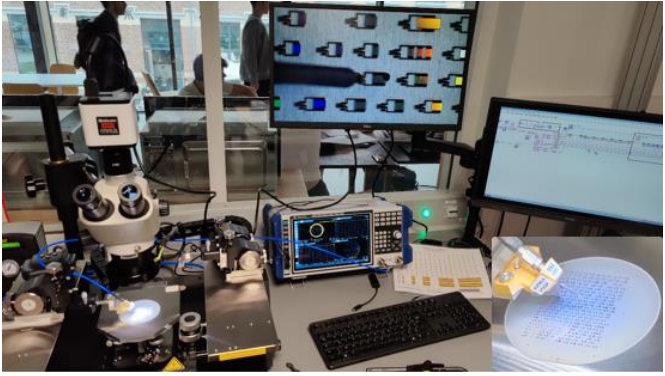

Fig.2 experimental setup used for electrical characterization, inset 4-inch ST-quartz substrate, 216 devices were characterized.

The electrical performances of fabricated SAW devices were evaluated based on their resonance frequency, quality factor, and impedance matching. Scattering parameters were measured using a Vector Network Analyzer from Rodh & Schwartz, which operates from 9 kHz to 6 GHz. The electrical parameters were determined by analyzing data from the S-parameter. Figure **2** illustrates the experimental setup for SAW devices electrical characterization, which employed Microtech Probe station and 250-μm pitch RF GSG probes.

## III. BUTTERWORTH-VAN DYKE EQUIVALENT CIRCUIT

A Butterworth-Van Dyke (BVD) circuit model was used as an equivalent circuit of SAW resonator **[25], [26]**, facilitating the extraction of relevant parameters as shown in Figure **3**.

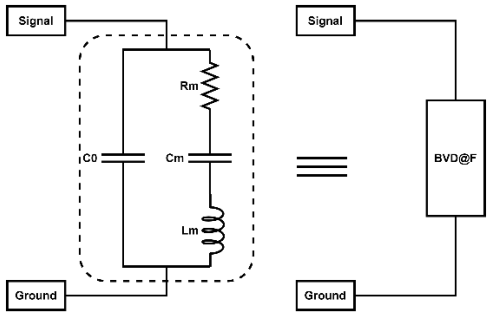

Fig.3 Butterworth-Van Dyke equivalent circuit.

The (BVD) model consists of a series resonator $L_m$ $C_m$ $R_m$ circuit, corresponding to the motional branch, coupled in parallel with a static capacitance $C_0$ which represents the fixed dielectric capacitance of the resonator. The admittance across such structure, shown in Fig. **3**, is given by Equation **2 [27]**, where the angular frequency $\omega = 2\pi f$.

$$Y(\omega) = \frac{R_m(\omega C_m)^2}{(R_m C_m \omega)^2 + (1 - L_m C_m \omega^2)^2} + j\left(\omega C_0 + \frac{C_m \omega - C_m^2 L_m \omega^3}{(C_m R_m \omega)^2 + (1 - C_m L_m \omega^2)}\right) \quad (2)$$

The admittance response of the SAW resonator including the resonance frequency and anti-resonance frequency can be obtained. At resonance, the imaginary part of admittance is zero, resulting in the following equations :

$$C_0 = \frac{C_m - C_m^2 L_m \omega^2}{(C_m R_m \omega)^2 + (1 - C_m L_m \omega^2)} \quad (3)$$

$$C_m = C_0 \left[\left(\frac{\omega_a}{\omega_r}\right)^2 - 1\right] \quad (4)$$

$$L_m = \frac{1}{\omega_r^2 C_m} \quad (5)$$

$$R_m = \frac{\omega_r L_m}{Q_r} \quad (6)$$

This results in two fundamental resonance frequencies corresponding to the maximum and minimum of the admittance amplitude, respectively.

$$f_r = \frac{1}{2\pi\sqrt{L_m C_m}} \quad , \quad f_a = f_r\sqrt{\frac{C_m + 1}{C_0}} \quad (4)$$

Where $f_r$ is the resonant and $f_a$ the anti-resonant frequencies.

The approach for calculating $Q_r$ and $Q_a$, quality factors at the resonance and anti-resonance respectively, relies on calculating the slope of the phase characteristic of the impedance characteristics.

The BVD model has been applied on two different developed designs and optimized to fit with a topology of multiplexed single chip SAW resonator, which will be discussed in section **IV**. Details of equivalent circuit code are given in the Appendix.

## IV. RESULTS AND DISCUSSION

### A. Single port resonator

An optimization in the fabrication process was conducted on different parameters: $N_p$, W, $N_r$, and $N_{br}$ to enhance the electrical response of the studied resonators. The design adjustments were limited by the fact that the IDT could not exceed 100 electrodes, constrained by the size of the exposure window, related to gas application that will be conducted later. The $S_{11}$ scattering parameter responses, depicted in Figure **4a** and **4b**, compare two fabricated resonators with quasi-identical design, except that one incorporated one Bragg reflector ($N_{br}$=1) and the other four ($N_{br}$=4).

Initial measurements aim to extract key resonator parameters, including the electromechanical coupling coefficient ($K^2$) and the quality factor (Q-factor). The objective was to assess whether the properties of the studied resonators align with the requirements for wireless interrogation. Effective electromechanical coupling coefficient values have then been calculated from the admittance measurements through the determination of the resonant and anti-resonant frequencies $f_r$ and $f_a$ **[28]**:

$$K^2 = \frac{\pi^2}{4} \frac{f_a - f_r}{f_a} \quad (5)$$

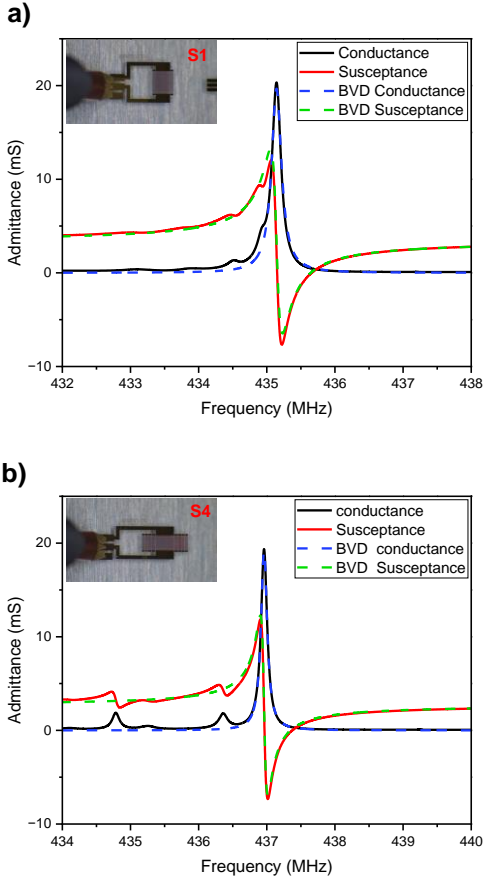

Fig.4 $S_{11}$ scattering Parameter, comparaison of admittance responses of resonator and BVD model for two designs incorporating a) two IDT reflectors and b) four IDT reflectors.

Devices exhibit distinct resonant frequencies $f_0$ near 434 MHz band (435.15 MHz and 436.96 MHz respectively), with a peak conductance Gres of 20.3 mS and 19.3 mS, respectively. At the resonant frequency, the $S_{11}$ gain parameter was measured to be -54.8 dB and -32.8 dB respectively. The Q-factor was calculated using the resonant frequency and the bandwidth $\Delta f$ and compared to the calculated one based on the BVD model. According to Equation 3, the Q-factor was estimated to be approximately 2466 for the single reflector device (S1) compared to 4003 for the four reflector's device (S4). The result shows that the Q-factor is enhanced as the number of reflectors increases.

$$Q = \frac{f_{res}}{\Delta f_{-3dB}} \quad (6)$$

Measured and BVD-fitted quality factor values are sufficiently close. The BVD-fitted electromechanical coupling coefficients $K^2$ for S1 and S4 designs are 0.30% and 0.24 % respectively. These values are quite similar to those described in literature for ST-Quartz made with heavy electrodes, which is ~0.3% [29]. These results are expected for such a structure. Extracted parameters of the two discussed designs are shown in table 2.

To investigate the behavior of our sensors, the measured impedance at the resonant frequency is *49.8 - 0.09j* Ohms for device S1 and *52.2 – 0.58j* for device S4, achieving good matching performances.

Table 2. BVD model parameters

| Parameters | S1 | S4 |
|---|---|---|
| $C_0$ | 1.232 pF | 0.974 pF |
| $C_m$ | 2.925 fF | 1.765 fF |
| $R_m$ | 50.824 Ω | 52.135 Ω |
| $L_m$ | 45.727 μH | 75.154 μH |
| $Q_{BVD}$ | 2460 | 3958 |
| $Q_{Measured}$ | 2466 | 4003 |
| $K^2$ | 0.30 % | 0.24 % |
| $K^2_{BVD}$ | 0.29 % | 0.22 % |
| $f_r$ | 435.183 MHz | 436.960 MHz |
| $f_a$ | 435.659 MHz | 437.365 MHz |

As mentioned before, our study was conducted on an ST-quartz wafer containing 216 uniformly distributed resonators. Electrical characterization leads to 143 out of 216 operational resonators, achieving a yield of 66%. The variations depicted in Figure 5 are attributable to the distributions of the impedance at the resonant frequency of each resonator design. The presented distributions results from the variations of the real and imaginary impedance values regarding the number of IDT pairs in reflector (Nr) and transducer IDT pairs (Np).

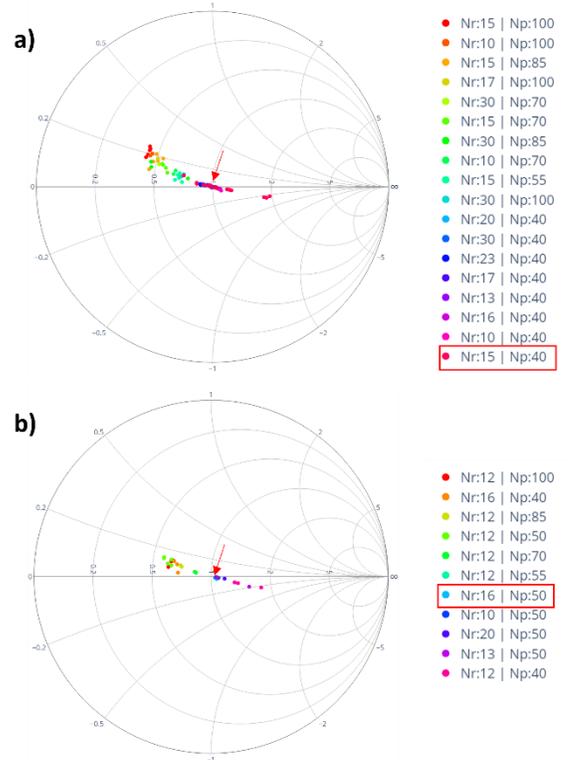

Figure 5 $S_{11}$ Smith Shart. Distribution of matching performances at the resonant frequency depending on the number of a) IDT pairs in reflector (Nr) and b) transducer IDT pairs (Np).

Most importantly, the combination of those two distributions results, at the resonant frequency, shows that an impedance matching 50 Ohms is achieved for a pair of parameters (Nr, Np) equal to (15,40) and (16,50), for the S1 and S4 resonator's configurations respectively. This configuration allows the system to act as a matched load, ensuring optimal power transfer from the antenna or cable to the resonator. For devices presenting impedance slightly above the target of 50 Ohms, a matching circuit could be employed to address any significant power reflection due to this mismatch.

Statistical analysis was also performed on the extracted data to determine whether the resonant frequencies of the produced devices is significantly different from the theory. Figure **6** illustrates a linear correlation between the resonant frequency and aperture. Changing the $N_{br}$ parameters results in a different base resonant frequency: 434.6 MHz and 436.3 MHz for S1 and S4 designs, respectively. Both designs exhibit an increase of 2% per W in resonant frequency.

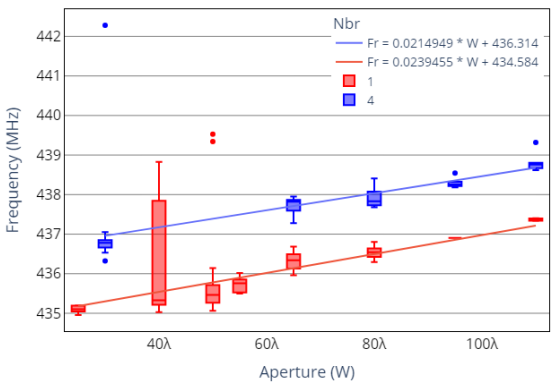

Figure 6 Resonant frequency with different aperture values, in red one IDT reflectors design, in blue four IDT reflectors design.

This can be explained by the fact that an increase in W leads to an increase in $C_0$ according to the Equation **7** [30],

$$C_0 = N_p(\varepsilon_r \varepsilon_0 + \varepsilon_0)W \tag{7}$$

Where $\varepsilon_r$ is the relative dielectric constant of the quartz substrate and $\varepsilon_0$ is the dielectric constant under vacuum.

This means that the electromechanical coupling becomes lower, indicating weaker coupling between the electrical and mechanical systems. When less energy is efficiently converted, it modifies the dynamic behavior of the resonator, which can lead to a change in the resonance condition.

Moreover, the intrinsic variation in the metallization ratio and the thickness of the metal layers used in the fabrication of these devices can also influence the resonant frequency. Variations in these parameters can lead to shifts in the resonant frequency, which may be used to tune specific frequency responses. Based on these dynamics, the design of multifrequency resonator devices operating in the 433MHz band may become a promising option for many applications [31]. By carefully manipulating the electrode width, metallization ratio, and metal thickness, it is possible to create devices capable of operating efficiently at multiple frequencies within the 433MHz band. This capability can be particularly beneficial in applications such as wireless communications, where devices often need to operate across different channels within the same frequency band.

## B. Multiplexed Single-port resonator

In addition to the two designs previously discussed, we propose an alternative approach that utilizes multifrequency sensors. Our approach involves arranging resonators sequentially, with each resonator tuned to its own distinct wavelength. On the other hand, this would result in a compact device that can be integrated for multi-gas detection systems. This device holds multiple working frequency that can be simultaneously read-out by a single reader antenna, ensuring that signals from different sensors do not interfere with each other.

Using the methodology introduced in Section **III**, multiplexed resonator parameters can be calculated by obtaining the BVD model of each single frequency. Measurements were then fitted based on a modified Butterworth Van Dycke model (Multi-MBVD) shown in Figure. **7a**. A shunt circuit, consisting of a resistance ($R_s$) and inductance ($L_s$) connected in parallel, was added to offset losses coming from parallel connection of each single device. The series circuit consisting of resistances ($R_c$) were used to simulate the wire connection between each single resonator.

Figure **S2** illustrates the iterative method for fitting measurements with the multi-MBVD model. Initially, the measured datas $Y_m$ are imported and the frequency range for the desired mode is set. Subsequently, the equivalent circuit parameters ($C_m$ $L_m$ $R_m$ and $C_0$) were fitted and resonator parameters ($f_r$, $f_a$, $Q_r$ and $K^2$) were determined based on a single device BVD equivalent circuit, according to equations 4 through 5. The subsequent iterative fitting process employs a nonlinear least square fitting method, referenced in **[32]**, with the goal of minimizing the difference between the measured admittance, $Y_m$, and the fitted admittance, $Y_f$. This was achieved by comparing both the real and imaginary components of admittance. The model was then extended to a multi-transducer network, from which all resonance peak parameters were extracted separately, ranging from the lowest ($f_0$) to the highest ($f_n$) frequency peak.

These BVDs models were then connected in parallel with both a shunt and cable modelisation. Where the resistance ($R_s$) and inductance ($L_s$) of the shunt and the resistance ($R_c$) of the cable model were determined. Corrections were then applied to the multi-MBVDs' resonance frequencies, shunt, and the cable parameters. Upon the convergence of the iterative process, the extraction concludes the key fitting results presented in table **3**.

According to the above designs, three multiplexed sensors M3, M3' and M5 were fabricated. M3 and M3' incorporate three resonators based on two (S1) and four (S4) reflectors respectively, while M5 have five-frequency response based on (S1) design. These implementations illustrate that multi-frequency devices can be effectively engineered by modifying the wavelength according to the proposed design concept, yielding multiple frequency responses. The Q-factor of the proposed devices was analyzed (Figure. **7b**), revealing a similar trend to that seen in single-frequency devices. Notably, the Q-factor is significantly higher in the M3' device confirming that an increasing number of reflectors enhances the electrical performance. Furthermore, all three designs showed a linear shape in the Q-factor, where the higher frequency devices and those closer to the RF probe present higher Q-factor.

The $S_{11}$ scattering parameter responses of the three multi-frequency SAW resonators are shown in figure **7c**, **7d** and **7e**.

The solid lines represent the measured responses compared to the dashed lines of the proposed muti-MBVD circuit where extracted parameters are provided in Table. **3**.

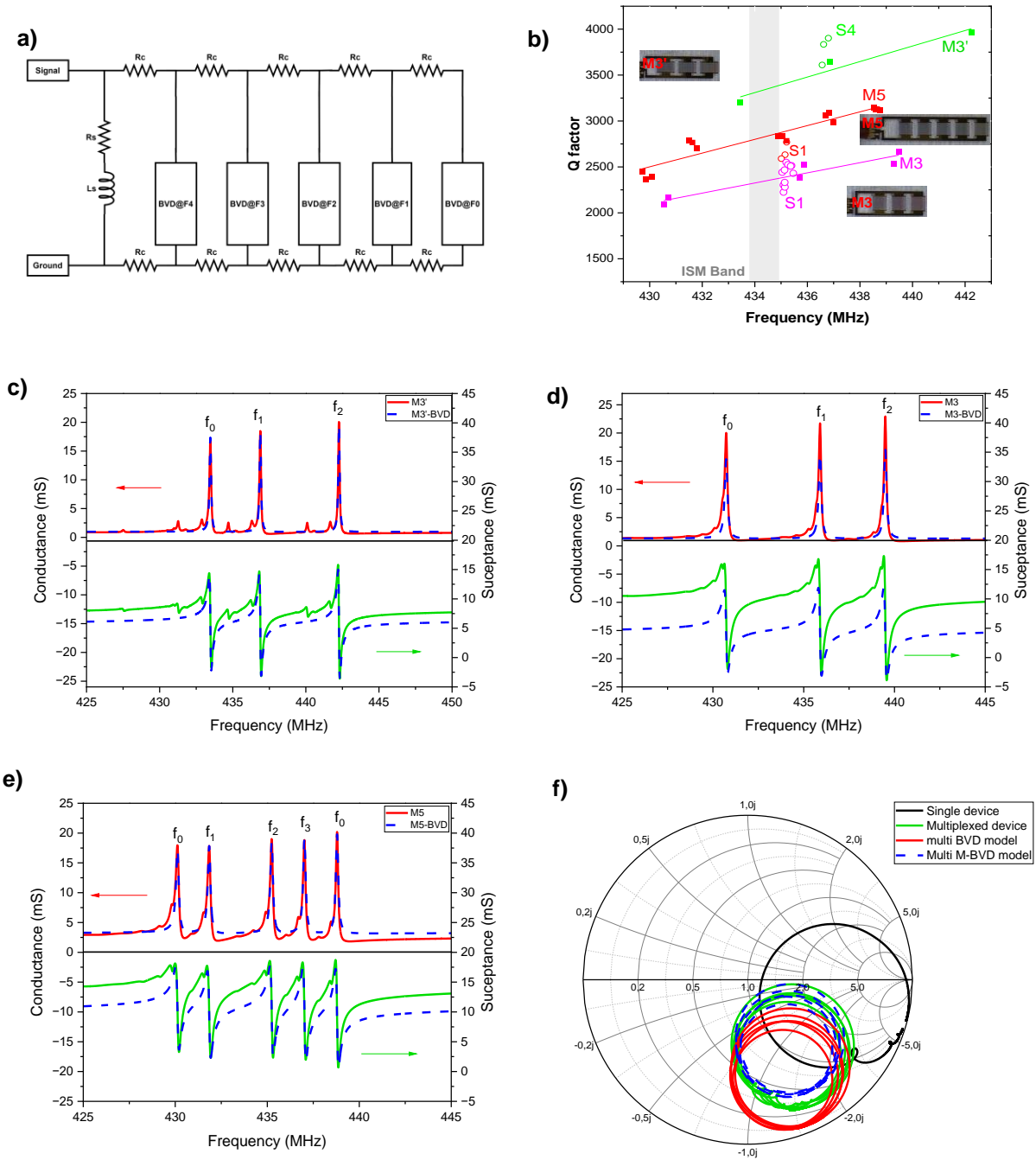

Figure 7 a) Equivalent circuit of the proposed multi-MBVD circuit b) Quality factor versus resonant frequency for three studied multiplexed devices (in green: M3', in red: M5 and in blue: M3). $S_{11}$ scattering Parameter, comparaison of admittance responses of resonator and M-BVD model for three multiplexed designs incorporating c) three working frequencies based on four IDT reflectors d) three working frequencies based two IDT reflectors and e) five working frequencies based two IDT reflectors. f) $S_{11}$ Smith Shart comparing the real response of a single device S1 (black line), the real response of a five frequencies multiplexed device M5, the multi BVD model for each seperated resonant peak (red lines) and the multi M-BVD model (dashed blue lines)

Table 3. Multi-MBVD model parameters

| Design | $C_0$ (pF) | $C_m$ (pF) | $R_m$ (Ω) | $L_m$ (μH) | $R_s$ (Ω) | $L_s$ (pF) | $R_c$ (pF) | Q | $Q_{M-BVD}$ | $K^2$ | $K^2_{m-BVD}$ | FOM | $FOM_{M-BVD}$ |
|---|---|---|---|---|---|---|---|---|---|---|---|---|---|
| *M3* | | | | | | | | | | | | | |
| $f_0$ | 1.012 | 2.23 | 62.513 | 58.795 | 74.727 | 88.262 | 1.0 | 2163.42 | 2413.40 | 0.101 | 0.147 | 218.235 | 382.613 |
| $f_1$ | 1.012 | 2.23 | 63.548 | 59.769 | 74.727 | 88.99 | 1.0 | 2527.49 | 2600.27 | 0.099 | 0.154 | 249.984 | 395.689 |
| $f_2$ | 1.012 | 2.23 | 65.083 | 61.212 | 74.727 | 90.058 | 1.0 | 2663.65 | 2652.38 | 0.111 | 0.18 | 295.272 | 458.14 |
| *M3'* | | | | | | | | | | | | | |
| $f_0$ | 0.97 | 1.782 | 52.704 | 72.69 | 189.458 | 158.208 | 1.023 | 3204.30 | 3478.01 | 0.076 | 0.09 | 244.815 | 345.817 |
| $f_1$ | 0.97 | 1.782 | 54.01 | 74.491 | 189.458 | 160.156 | 1.023 | 3646.63 | 3732.64 | 0.08 | 0.097 | 291.49 | 365.829 |
| $f_2$ | 0.97 | 1.782 | 54.864 | 75.67 | 189.458 | 161.418 | 1.023 | 3968.6 | 3838.98 | 0.082 | 0.103 | 325.875 | 386.013 |
| *M5* | | | | | | | | | | | | | |
| $f_0$ | 1.014 | 2.218 | 59.047 | 59.308 | 145.637 | 68.462 | 1.002 | 2394.81 | 2523.54 | 0.054 | 0.059 | 129.494 | 162.198 |
| $f_1$ | 1.014 | 2.218 | 59.526 | 59.789 | 145.637 | 68.739 | 1.002 | 2704.07 | 2770.59 | 0.053 | 0.061 | 142.319 | 169.182 |
| $f_2$ | 1.014 | 2.218 | 60.014 | 60.28 | 145.637 | 69.021 | 1.002 | 2785.14 | 2830.85 | 0.053 | 0.061 | 148.432 | 168.775 |
| $f_3$ | 1.014 | 2.218 | 60.961 | 61.231 | 145.637 | 69.563 | 1.002 | 2984.40 | 2953.14 | 0.054 | 0.065 | 162.185 | 178.108 |
| $f_4$ | 1.014 | 2.218 | 61.451 | 61.723 | 145.637 | 69.842 | 1.002 | 3122.76 | 2924.36 | 0.063 | 0.077 | 196.242 | 208.706 |

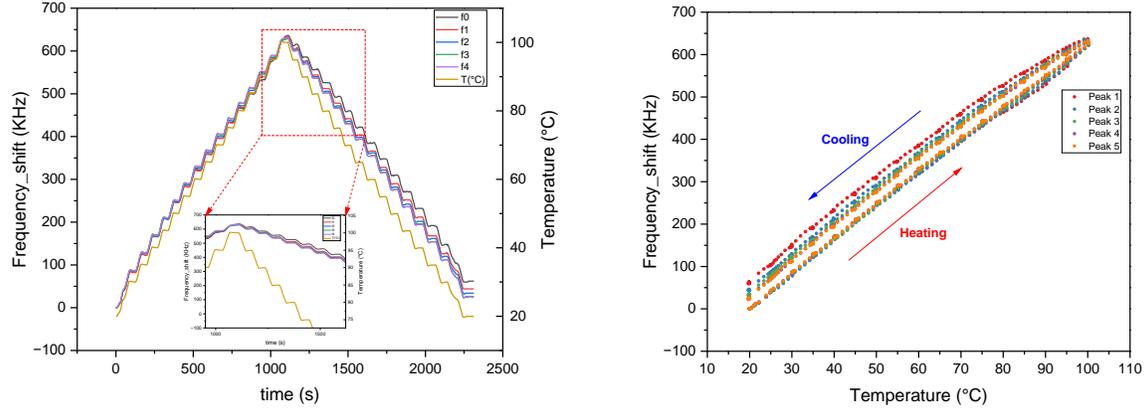

Figure 8 a) Resonant frequencies shift of the multifrequency single-chip SAW resonator versus temperature in a heating and cooling cycle b) linear evolution of the frequency response of each resonant frequency from room temperature to 100°C.

The experimental $S_{11}$ curve of the resonator M5, in which five resonant frequencies corresponding to the different studied wavelengths, i.e., $f_0$=438.82 MHz, $f_1$=437.04 MHz $f_2$=435.25 MHz, $f_3$=431.88 MHz, and $f_4$=430.15 MHz are observed. Several features are noteworthy in the electrical response of this design. The maximum gain for each sensor is approximately the same. In contrast the minimum loss for each of the five resonant frequencies increase monotonically with the increase of the frequency, a result of frequency-dependent coupling of electromagnetic energy between input and output transducers.

To understand the physical mechanism, the $S_{11}$ parameter is discussed in a Smith chart. Figure **7f** shows that the $S_{11}$ response of the measured SAW resonators (green line); it is not matched with the proposed (Multi-BVD) (red line), which means that each individual filter in the optimized multiplexed resonator is not suitable for use in its frequency band. However, if we connect our model to a shunt by adding $R_c$ resistance (blue dashed line), the new (multi M-BVD) circuit modelling based on the structure shown in Fig. **7a** is in good agreement with the measured results if some spurious waves are ignored.

*C. Temperature Measurments*

A temperature reading was demonstrated by measuring S-parameters of the single chip multi-frequency device with a vector network analyzer (VNA) at different positions. The measurements were conducted within a micro-probe system from the Nextron company by applying various temperature levels between room temperature and 100 °C (see setup in Figure **S3**). The data was analyzed with dedicated analysis scripts (python script in the supporting information) to retrieve the relevant sensor data, like frequency shifts, quality factor, the peak and signal quality, the temperature coefficient of frequency (TCF) and the effective electromechanical coupling coefficient $K^2$. SAW sensors typically exhibit sensitivity to temperature, and this sensitivity is usually assessed through the temperature coefficient of frequency (TCF). For ST-cut quartz crystal, the TCF is known to be nearly zero. Figure **8a** shows the frequency versus temperature characteristics, It can be seen that the five resonant frequencies exhibit a step-up and step-down change when an increase and a decrease of temperature is done in the range of 20 to 110 °C. The ST-cut quartz crystal provides a linear response to temperature variation, and the frequency-temperature linearity is reproducible (Figure 8b). This reproducibility demonstrates that gold electrodes maintain their stability throughout the characterization process. The TCF of SAW sensor is calculated to be ∼18 ppm/°C, indicating that the ST-cut quartz substrate is not sensitive to temperature.

The effective electromechanical coupling coefficient, $K^2$, remains stable across the entire investigated temperature range, with values consistently close to 1.4% ± 0.2% (refer to Figure.S4). Minor fluctuations observed in $K^2$ are primarily attributed to uncertainties in reading the parallel and series resonance frequencies from the admittance curves. These uncertainties arise due to the presence of small spurious modes, as the variations are not entirely reproducible from one device to another. This stability is highly promising for the future development of reliable temperature under human breath measurements condition.

## V. CONCLUSION

This paper discusses a novel approach in the development of multiplexed one-port saw resonators, fabricated on a single chip using 42.75° Y (ST) cut Quartz. The fabrication process focused on critical parameters such as aperture size, the number of interdigitated electrode (IDT) pairs, IDT wavelengths, and number of reflectors. These parameters were experimentally optimized to enhance electrical responses of the proposed devices. Relevant parameters such as resonance frequency, quality factor and electromechanical coupling were calculated and compared with extracted ones through a modified Butterworth-Van Dyke (Multi-MBVD) model. The experimental results closely aligned with the simulation results, validating the model's efficacy. Finally, temperature measurements were demonstrated by measuring S-parameters of the single chip multi-frequency device. The results suggest that these multiplexed designs hold potential for applications such as electronic noses for gas detection.

## ACKNOWLEDGMENT

**Author Contributions:** B.H. conceived of the project. B.H. and A.W. designed and conducted the experiments. O.S, B.H and A.W conducted the theoretical study for the BVD model. A.W. conducted M-BVD equivalent circuit simulations, characterization and datas analysis. P.D and M.L conducted the Nextron setup for temperature measurements. B.H. and AW wrote the paper. All authors have read and agreed to publish this version of the manuscript.

**Funding:** The authors gratefully acknowledge financial support from the Hauts-de-France region, the CNRS' through the "COSMOS2" STIMulE project and the French National Nanofabrication Network RENATECH for micro and nanofabrication.

Special thanks to Pascal Delmotte and Saliha Wendi for their help during the fabrication process.

**Conflicts of Interest:** The authors declare no conflict of interest.

**Data Availability:** All data supporting this study are openly available in a GitHub link:

[27] J. R. Lane, " integrating superconducting qubits with quantum fluids and surface acoustic wave devices ", Ph.D. dissertation, Dept. of Physics, Michigan State Univ., East Lansing, MI, USA, 2022.

[28] *IEEE Standard on Piezoelectricity,"* in ANSI/IEEE Std 176-1987 , vol., no., pp.0_1-, 1988.

[29] M. Kadota, " Resonator Filters Using Shear Horizontal-Type Leaky Surface Acoustic Wave Consisting of Heavy-Metal Electrode and Quartz Substrate ", vol. 51, n° 2, 2004.

[30] Y. Liu *et al.*, " High-Performance SAW Resonator with Spurious Mode Suppression Using Hexagonal Weighted Electrode Structure ", *Sensors*, vol. 23, n° 24, p. 9895, déc. 2023.

[31] Malarvezhi, P., Dayana, R., Subhash Chandra, T.V.S, "A Comprehensive Survey on RF Module at 433 MHz for Different Applications", In: Hemanth, D., Vadivu, G., Sangeetha, M., Balas, V. (eds) Artificial Intelligence Techniques for Advanced Computing Applications. Lecture Notes in Networks and Systems, vol 130. Springer, Singapore, 2021.

[32] D. M. Bates et D. G. Watts, "*Nonlinear regression analysis and its applications"*. in Wiley series in probability and mathematical statistics, Applied probability and statistics, New York: Wiley, 1988.

## AUTHORS

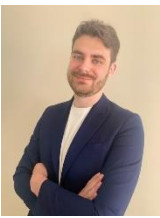

**Alexandre WESTRELIN** recently graduated from the ICAM engineering school in Lille. He holds an engineering degree specializing in the fields of transport, logistics, and energy. During his studies in 2022, Alexandre actively participated in a research memoir focused on applied material science. His research involved the design and development of an integrated platform for the characterization of SAWs (Surface Acoustic Wave) sensors. This project showcased his expertise and interest in cutting-edge technologies. Currently, Alexandre is employed at "MC2 Technologies," a prominent company based in Villeneuve d'Ascq. The company specializes in the development of advanced RF microwave-based protection systems. Within MC2 Technologies, Alexandre serves as an automation engineer, responsible for system installation and testing. His role involves ensuring the smooth integration and functionality of the company's cutting-edge systems.

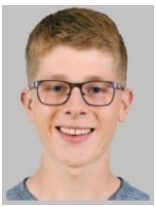

**Pierre Debavelear** is an apprentice student at ICAM school engineering ICAM, Lille, France. As part of his final year, he conducted a six-month research project between September 2024 and mid-February 2025, under the supervision of Dr. Bilel Hafsi. This project focused on the characterization of SAW sensors for gas detection. Pierre is currently working at SNCF (French National Railway Company) as part of the department focused on the evolution and modernization of test systems. His work involves the development and integration of innovative testing methods and tools to support maintenance and reliability operations within the railway sector.

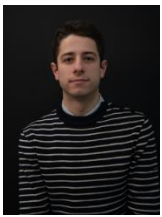

**Mathieu Lefebvre** is an apprentice student at ICAM school engineering, Lille, France. As part of his final year, he conducted a six-month research project between September 2024 and mid-February 2025, under the supervision of Dr. Bilel Hafsi. This project focused on the characterization of SAW sensors for gas detection. The main objective of the project was the installation and implementation of a test bench enabling precise evaluation of sensor performance under controlled environmental conditions. In parallel with this project, Mathieu Lefebvre is currently working as a project manager apprentice at Framatome a French multinational company specializing in the design, construction, and maintenance of nuclear power plants and components.

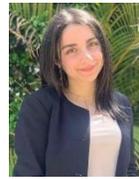

**Nour Abdallah** is currently a Master 2 student in Physics at the University of Lille, France. She is completing her master's internship at ICAM School of Engineering in Lille, working on the SNIFFER project. This research focuses on the development of a connected and autonomous sensing platform based on Surface Acoustic Wave (SAW) sensors for multivariate detection of complex gas mixtures. Her interests include sensor technology, gas detection systems, and embedded electronics for environmental and health monitoring applications.

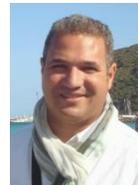

**Pr. Kamal Lmimouni** received the Degree in electronics, the Ph.D. degree, and the Research Direction Diploma from the University of Lille, Lille, France, in 1997 and 2006, respectively ,he was appointed as an Assistant Professor with the University of Lille in 2001, where he focused on charge transport in organic materials and design of new organic field-effect transistor structures including memory devices. He is a currently a Full Professor at the University of Lille.

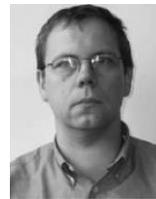

**Dr. Olivier Stienne** received the engineering degree in electronics engineering from the Cnam Engineering School, Paris, France, in 1999. Obtained his Ph.D. degree in micro and nanotechnologies, acoustic, and telecommunications within the Interference and Electromagnetic Compatibility Group, Lille University, Villeneuve d'Ascq, France. He worked for 10 years in the industry. Then, he joined the E.E.A. Department, Icam, Lille, where he teaches electronics and telecommunications to engineering students. His current research interests include the analysis of the impact of electromagnetic interference on the performance of wireless communications using orthogonal frequency division multiplexing modulation.

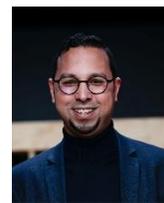

**Dr. Bilel HAFSI** received the master's degree in materials, nanostructures, devices et microelectronics systems from Monastir University, Tunisia, in 2012 and the PhD. degree in micro et nanoelectronics, acoustic and telecommunications from Lile University, France, in 2016. From 2016 to 2018, he was a Teacher and Research Assistant with the central Lille institute. His research interest includes the development of Devices based organic materials, and fabrication of micro- or nanostructured surfaces. To expand these concepts, he focused on developing new kind of memory devices based on double floating gate made of gold nanoparticles and reduced graphene oxide. his main interest was to boost the memory performance (endurance, memory window and retention time). From 2018 to 2019 he has experienced the Job of a Dream consultant in private company "Techshop Les Ateliers Leroy Merlin" where he supervised many projects in the field of IOT's, Robotics ...Currently, he works as associate professor in the electrical department in ICAM school of engineering, Lille 59800, France. Dr. Bilel HAFSI was a recipient of the International research Young Scientist Award in 2018 from the doctoral college in the north of France.